\documentstyle[prb,aps]{revtex}
\begin{document}
\preprint{UCF-CM-97-002}
%
%
\title{Ensemble Density Functional Approach to Charge-Spin 
Textures in Inhomogeneous Quantum
Hall Systems}
\author{O. Heinonen}
\address{Department of Physics, University of Central Florida, FL 32816-2385}
\author{J.M. Kinaret}
\address{Department of Applied Physics, G\"oteborg University, and\\
Chalmers University of Technology,
S-412 96 G\"oteborg, Sweden}
\author{M.D. Johnson}
\address{Department of Physics, University of Central Florida, FL 32816-2385}
\maketitle
\begin{abstract}
We extend our ensemble density functional approach to quantum Hall systems to
include non-collinear spins
to study charge-spin
textures in inhomogeneous quantum Hall systems. We have 
studied the edge reconstruction in quantum dots at unit bulk
filling factor and at 1/3 bulk filling factor as a function of the
stiffness of an external confining potential.
For soft enough edges, these systems reconstruct to a state in which the electron
spins rotate gently as the edge is approached, with a non-trivial spin-charge
texture at the edge of the system.
\end{abstract}
\pagebreak
\section{Introduction}
\label{intro}
It has recently become clear that the spin degree of 
freedom plays a significant
role near ferromagnetic fillings in the quantum Hall 
effect (QHE)\cite{Sondhi,Fertig,Girvin_MacDonald_review}. This is because
of the low ratio of the Zeeman 
energy $E_Z=g^*\mu_B B$ to the Coulomb energy $E_C=e^2/(\epsilon_0 l_B)$. 
Here, $g^*$ is the
Land\'e $g$-factor, $\mu_B$ the Bohr magneton, $B$ the applied magnetic
field strength,
$\epsilon_0$ the static dielectric constant, 
and $l_B=\sqrt{\hbar c/eB}$ the
magnetic length. For GaAs systems, the low values of the Land\'e factor 
and of the
electron effective mass $m^*$ conspire to make 
$\tilde g=E_Z/E_C\approx E_Z/(\hbar\omega_c)\alt0.02$ for 
fields in the range
of a few tesla to about ten tesla. Here, $\omega_c=eB/(m^* c)$ is the cyclotron
frequency. Nonetheless, single-particle spin-flip excitations still cost
a large amount of energy, because of the loss of exchange energy associated with a
spin flip. This means that the spin degree of freedom is controlled by the
inter-electron Coulomb energy, and not by the Zeeman energy. One consequence is that
if a single spin is reversed, it becomes
energetically favorable for the system to
smoothly rotate the magnetization direction to restore it over some distance
from the reversed spin. Due to the connection
between flux and charge density in incompressible
ferromagnetic QHE ground states, 
such spin textures also acquire
a charge density, and the resulting spin-charge textures are 
commonly called `skyrmions'. (More accurately, skyrmions are the particular
type of spin-charge textures that show up in certain continuum models, such
as the nonlinear $\sigma$-model\cite{Skyrme}.)
There is now ample theoretical\cite{Fertig,Brey1,Kamilla} and experimental 
work\cite{Barrett,Schmeller,Aifer} suggesting that such
skyrmions are indeed the low-energy charged excitations, 
at least near filling
factor $\nu=1$. This is, for example, manifested in the rapid 
destruction of the ground state
polarization observed experimentally\cite{Barrett} 
as the filling factor is varied away from
unity.
Recent theoretical work\cite{Karlhede1,Franco,Karlhede2,Oaknin} 
has also indicated that the edge reconstruction of
ferromagnetic QHE systems may acquire non-trivial spin textures (or charge
density waves) as the edge confinement is softened. This may radically
alter our understanding of edges in QHE systems, and, concomitantly, 
our interpretations of experiments which probe the gapless edge modes.
This is at the present best understood for edges of $\nu=1$ systems.
One question we wish to address here is whether similar charge-spin
textures can occur at the edges of systems in regimes of the ferromagnetic fractional
quantum Hall effect (FQHE) (such as $\nu=1/3$), where not just electron exchange
but also correlations are important.

We have previously developed an 
ensemble density functional approach
for spin-polarized systems\cite{Heinonen1,Heinonen2}, and subsequently generalized 
that approach to include
the electron spin, but with the spin quantization axis constrained to be parallel
to the external magnetic field\cite{Lubin}. We present here a further 
extension which
is able to deal with a rotating spin quantization axis. 
Advantages of
our DFT approach are that it includes
electron interactions beyond exchange, and that it can be applied to large
inhomogeneous fractional QHE systems. 
This makes our ensemble spin DFT approach
the only available method which can be applied to general inhomogeneous
QHE systems, spanning regimes from the FQHE to the semiclassical, and which includes
the spin degree of freedom and Landau-level mixing.
We have used this approach to study the edge reconstruction
of circularly symmetric quantum dots. Our
results show that as the edge confinement is softened, the system goes from
a spin-polarized sharp edge to a softer edge 
with a non-trivial spin texture, in
agreement with results obtained by other groups using the 
Hartree-Fock approximation (at integer filling) 
or field-theoretical models\cite{Karlhede1}. 
A new result here is to show how the reconstruction to a spin-charge textured edge
can also happen for FQHE systems.

This paper is organized as follows. In Section \ref{spins}, we present the
general ensemble density functional theory for non-collinear spins. Section
\ref{technical} presents some technical details of the theory, including an
extension of previously used exchange-correlation energies for QHE systems to
include both higher Landau levels and electron spin. In Section \ref{dot} we
present results from numerical calculations of quantum dots at unit bulk
filling factor 
and at 1/3 bulk filling factor. Finally, section
\ref{end} contains conclusions and a discussion.

\section{Ensemble spin density functional theory for non-collinear spins}
\label{spins}
In its most general formulation, spin density functional
theory, as developed by von Barth and Hedin\cite{von_Barth}, allows for non-collinear
spins. This is based on a generalization
of the Hohenberg-Kohn\cite{Hohenberg} theorem in which the electron density $n({\bf r})$
is replaced by the single-particle density matrix
$\rho_{\sigma,\sigma'}({\bf r})\equiv\langle 0|\hat\psi^\dagger_\sigma({\bf r})
\hat\psi_{\sigma'}({\bf r})|0\rangle$, where $\hat\psi_{\sigma}({\bf r})$
($\hat\psi^\dagger_\sigma({\bf r})$) is the usual electron annihilation
(creation) operator for an electron of spin $\sigma$ at
position ${\bf r}$, and $|0\rangle$ is the ground state. We will use notations
in which $\sigma=\pm1$ or $\sigma=\uparrow,\downarrow$, with up-spin ($\uparrow$)
corresponding to $\sigma=-1$. A computationally
useful approach is then constructed in the usual way by considering
an auxiliary non-interacting system in some effective external
potential $v_s({\bf r})$ chosen so that this system has
the same ground state single-particle density matrix $\rho_{\sigma,\sigma'}({\bf r})$
as the interacting system at hand. A variational principle
associated with the generalized Hohenberg-Kohn theorem then yields
Kohn-Sham (KS) equations\cite{Kohn_Sham}, 
which now include spin-dependent exchange-correlation
potentials 
\begin{equation}
V_{{\rm xc},\sigma\sigma'}({\bf r})\equiv
{\delta E_{\rm xc}\left[\rho_{\sigma,\sigma'}({\bf r})\right]
\over 
\delta \rho_{\sigma,\sigma'}({\bf r})}.
\label{eq:spins:V_xc_sigma}
\end{equation}
A difficulty is that one does not usually have
reliable approximations for the exchange-correlation potentials
$V_{{\rm xc},\sigma\sigma'}{(\bf r})$, not even in the local density
approximation (LDA). Consequently, the density matrix is usually assumed to be
diagonal for all ${\bf r}$, 
which means that the direction of the magnetization is assumed to be
constant, and that direction is then conveniently chosen as
the spin quantization axis. Thus, only up- and down-spin densities
enter into the KS equations, and for the LDA (or extensions including 
generalized gradient approximations), one
only needs to know the exchange-correlation energy $E_{\rm xc}(n,\xi)$
of a uniform system of density $n=n_\uparrow+n_\downarrow$ and
polarization $\xi\equiv(n_\uparrow+n_\downarrow)/(n_\uparrow-n_\downarrow)$.
There exist now very accurate calculations of $E_{\rm xc}(n,\xi)$ 
for two- and three-dimensional electron gases (in zero
magnetic field)\cite{Tanatar,Perdew}.

However, the approximation of constant magnetization direction
obviously does not work in systems for
which it is known that the magnetization direction changes in space. Examples
of such systems are Mn$_3$Sn\cite{Sticht}, 
$\gamma$-Fe\cite{Uhl}, U$_3$Pt$_4$\cite{Sandratskii}, and 
QHE systems near unit filling. For
such systems the full single-particle density matrix has to be used, and
the problem then arises as to how one should construct a LDA. We
will here follow an approach developed by K\"ubler, Sticht and
co-workers\cite{Kuebler,Sticht}, and extend their approach to an ensemble DFT appropriate
for QHE systems. The basic idea is to locally rotate the spin
quantization axis to obtain a representation which locally diagonalizes the
single-particle density matrix. The advantage of this procedure is
that in order to construct
a LDA, one then only needs the exchange-correlation energy as a function
of spin-up and spin-down densities (or total density and polarization),
for which approximations exist.
The price one has to pay is to introduce local spin rotation angles
$\theta({\bf r})$ and $\varphi({\bf r})$, which complicates 
the KS equations. This approach has
given good results in applications to materials such as 
$\gamma$-Fe\cite{Uhl}, U$_3$Pt$_4$\cite{Sandratskii},
and $\alpha$-Fe$_2$O$_3$\cite{Sandratskii2}.
It can also give the spin stiffness important in studying spin-charge 
textures in the QHE. That this can happen in the LDA is not immediately
obvious -- in field-theoretical approaches the spin stiffness enters via a
gradient of magnetization, and such gradients are neglected in the LDA. We will
show in section \ref{technical} that the LDA does give a spin stiffness, although
formally its origin appears a bit different.

We now proceed to review the LDA approach of K\"ubler, Sticht and 
co-workers\cite{Kuebler,Sticht}.
We begin by writing
the ground state energy as a functional of the single-particle density matrix
for a two-dimensional system in a constant external magnetic field ${\bf B}=B\hat z$:
\begin{eqnarray}
E\left[\rho_{\sigma,\sigma'}({\bf r})\right]
 & = & T_s[\rho_{\sigma,\sigma'}({\bf r})]+
g^*\mu_B B\sum_{\sigma,\sigma'}\sigma\delta_{\sigma,\sigma'}
\int \rho_{\sigma,\sigma'}({\bf r})\, d^2r\nonumber \\
&&+\frac{1}{2}\int\int\,n({\bf r})v_H({\bf r}-{\bf r}')n({\bf r}')d^2r\,d^2r'
+E_{\rm xc}[\rho_{\sigma,\sigma'}({\bf r})].
\label{eq:spins:Exc1}
\end{eqnarray}
Here, $T_s[\rho_{\sigma,\sigma'}({\bf r})]$ is the kinetic energy functional
of non-interacting electrons, which in
our case includes the external
magnetic field ${\bf B}$. The particle density is 
$n({\bf r})={\rm Tr}\rho_{\sigma,\sigma'}({\bf r})$, $v_H({\bf r})$ is the
Hartree potential,
and $E_{\rm xc}\left[\rho_{\sigma,\sigma'}({\bf r})\right]$ is
the exchange-correlation energy, which depends parametrically
on the magnetic field (for ease of notation we omit 
this parametric dependence). At each point ${\bf r}$ we calculate an $SU(2)$
matrix ${\bf U}({\bf r})$ which locally diagonalizes $\rho_{\sigma,\sigma'}({\bf r})$:
\begin{equation}
\sum_{\sigma,\sigma'}U_{\alpha,\sigma}({\bf r})\rho_{\sigma,\sigma'}({\bf r})
U^*_{\sigma',\beta}({\bf r})=\delta_{\alpha,\beta}n_{\alpha}({\bf r}),
\label{eq:spins:SU2}
\end{equation}
with $n_{\alpha}({\bf r})$ the eigenvalues of $\rho_{\sigma,\sigma'}({\bf r})$.
We write ${\bf U}({\bf r})$ in the standard form
\begin{equation}
{\bf U}({\bf r})=\left(
\begin{array}{cc}
 e^{\frac{i}{2}\varphi({\bf r})}\cos\frac{\theta({\bf r})}{2} &
e^{-\frac{i}{2}\varphi({\bf r})}\sin\frac{\theta({\bf r})}{2} \\
-e^{\frac{i}{2}\varphi({\bf r})}\sin\frac{\theta({\bf r})}{2} &
e^{-\frac{i}{2}\varphi({\bf r})}\cos\frac{\theta({\bf r})}{2}
\end{array}
\right),
\label{eq:spins:SU2_matrix}
\end{equation}
where $\varphi({\bf r})$ and $\theta({\bf r})$ are the local azimuthal and
polar angles of the
magnetization density relative to a fixed coordinate system.
The requirement that ${\bf U}({\bf r})$ diagonalizes $\rho_{\sigma,\sigma'}({\bf r})$
then yields
\begin{equation}
\tan\varphi({\bf r})=-{{\rm Im}\rho_{\uparrow,\downarrow}({\bf r})
\over{\rm Re}\rho_{\uparrow,\downarrow}({\bf r})}
\label{eq:spins:tanphi}
\end{equation}
and
\begin{equation}
\tan\theta({\bf r})={
2\left[\left({\rm Re}\rho_{\uparrow,\downarrow}({\bf r})\right)^2+
\left({\rm Im}\rho_{\uparrow,\downarrow}({\bf r})\right)^2
\right]^{1/2}
\over
\left[\rho_{\uparrow,\uparrow}({\bf r})-\rho_{\downarrow,\downarrow}({\bf r})\right]},
\label{eq:spins:tantheta}
\end{equation}
with ${\rm Re}(z)$ and ${\rm Im}(z)$ denoting the real and 
imaginary parts of a complex
number $z$.
Equation (\ref{eq:spins:SU2}) gives
\begin{eqnarray}
n_{\uparrow}({\bf r}) & = & \rho_{\uparrow,\uparrow}\cos^2\frac{\theta({\bf r})}{2}
+\frac{1}{2}\rho_{\uparrow,\downarrow}({\bf r})e^{i\varphi({\bf r})}\sin\theta({\bf r})
+\frac{1}{2}\rho_{\uparrow,\downarrow}^*({\bf r})e^{-i\varphi({\bf r})}\sin\theta({\bf r})
+\rho_{\downarrow,\downarrow}\sin^2\frac{\theta({\bf r})}{2} \nonumber \\
n_{\downarrow}({\bf r}) & = &\rho_{\uparrow,\uparrow}\sin^2\frac{\theta({\bf r})}{2}
-\frac{1}{2}\rho_{\uparrow,\downarrow}({\bf r})e^{i\varphi({\bf r})}\sin\theta({\bf r})
-\frac{1}{2}\rho_{\uparrow,\downarrow}^*({\bf r})e^{-i\varphi({\bf r})}\sin\theta({\bf r})
+\rho_{\downarrow,\downarrow}\cos^2\frac{\theta({\bf r})}{2}.
\label{eq:spins:n_diag}
\end{eqnarray}

We now have
a representation in which 
$\rho_{\sigma,\sigma'}({\bf r})$ is locally diagonal, so that in the 
LDA we only need
to know the exchange-correlation energy $E_{\rm xc}(\nu,\xi)$ as a function
of total density and polarization, or, equivalently, $n_\uparrow$ and
$n_\downarrow$. By using the $SU(2)$ transformation, and by expressing the
single-particle density matrix in terms of occupied KS orbitals
$\psi_{i,\sigma}({\bf r})$ we can then write
the KS equations in the following form.
\begin{eqnarray}
T_s\psi_{i,\uparrow}({\bf r})+v_0({\bf r})\psi_{i,\uparrow}({\bf r})
+\Delta v({\bf r})\left[
\cos\theta({\bf r})\psi_{i,\uparrow}({\bf r})+
\sin\theta({\bf r})e^{i\varphi({\bf r})}\psi_{i,\downarrow}({\bf r})\right]
& = & \epsilon_{i,\uparrow}\psi_{i,\uparrow}\nonumber\\
T_s\psi_{i,\downarrow}({\bf r})+v_0({\bf r})\psi_{i,\downarrow}({\bf r})
+\Delta v({\bf r})\left[
\sin\theta({\bf r})e^{-i\varphi({\bf r})}\psi_{i,\uparrow}({\bf r})-
\cos\theta({\bf r})\psi_{i,\downarrow}({\bf r})\right]
& = & \epsilon_{i,\downarrow}\psi_{i,\downarrow},
\label{eq:spins:kuebler1}
\end{eqnarray}
for each single-particle spin doublet 
$\psi_i({\bf r})=\left(\psi_{i,\uparrow}({\bf r}),
\psi_{i,\downarrow}({\bf r})\right)$.
Here, 
\begin{equation}
v_0({\bf r})  =  v_{\rm ext}({\bf r})+v_{\rm H}({\bf r})+\frac{1}{2}\left[
v_{{\rm xc},\uparrow}({\bf r})+v_{{\rm xc},\downarrow}({\bf r})\right],
\end{equation}
and
\begin{equation}
\Delta v({\bf r})=\frac{1}{2}\left[v_{{\rm xc},\uparrow}({\bf r})
-v_{{\rm xc},\downarrow}({\bf r})\right],
\label{eq:spins:Delta_v}
\end{equation}
with
\begin{equation}
v_{{\rm xc},\sigma}=
{\delta E_{\rm xc}[n_\uparrow({\bf r}),n_\downarrow({\bf r})]
\over\delta n_{\sigma}({\bf r})}.
\label{eq:spins:v_xc_sigma}
\end{equation}
In the KS equations ({\ref{eq:spins:kuebler1}) there is a coupling between up- and
down-spin components, so that in general the KS orbitals are now two-component
spinors with both up- and down-spin components. Note that in an 
angular momentum representation, the equations ({\ref{eq:spins:kuebler1}) 
couple $z$-components of orbital angular momentum $L_z+m\hbar$, 
with $m$ an integer, and 
spin angular momentum $S_z=\frac{1}{2}\hbar$ to $(L_z,S_z=-\frac{1}{2}\hbar)$. 
This coupling will also provide the mechanism
for spin-charge textures in QHE systems in the same way as the Hartree-Fock
equations by Fertig {\em et al.} do\cite{Fertig}.
In the LDA, we write
\begin{equation}
E_{\rm xc}=\int\, d^2r\,n({\bf r})\epsilon_{\rm xc}[\nu({\bf r}),\xi({\bf r})],
\end{equation}
where $\epsilon_{\rm xc}[\nu,\xi]$ is the exchange-correlation
energy per particle in an infinite, homogeneous system of filling factor
$\nu$ and polarization $\xi=(\nu_\uparrow-\nu_\downarrow)/n$, and 
$\nu({\bf r})=2\pi l_B^2n({\bf r})$ is the density expressed as local
filling factor. Then
\begin{eqnarray}
v_{{\rm xc},\uparrow}({\bf r}) & = &
\left[{\partial\over\partial\nu}+{1\over\nu}\left(1-\xi\right){\partial
\over\partial\xi}\right]\left[\nu\epsilon_{\rm xc}(\nu,\xi)\right]
\nonumber\\
v_{{\rm xc},\downarrow}({\bf r}) & = &
\left[{\partial\over\partial\nu}-{1\over\nu}\left(1+\xi\right){\partial
\over\partial\xi}\right]\left[\nu\epsilon_{\rm xc}(\nu,\xi)\right],
\label{eq:spins:v_updown}
\end{eqnarray}
with the derivatives evaluated at $\nu=\nu({\bf r})$ and $\xi=\xi({\bf r})$,
so that
\begin{eqnarray}
v_0({\bf r}) & = &  v_{\rm ext}({\bf r})+v_{\rm H}({\bf r})+{\partial\over\partial\nu}\left[
\nu\epsilon_{\rm xc}(\nu,\xi)\right]-\xi{\partial\over\partial\xi}
\epsilon_{\rm xc}(\nu,\xi),\nonumber\\
\Delta v({\bf r})& = & {\partial\over\partial\xi}\epsilon_{\rm xc}(\nu,\xi).
\label{eq:spins:v_0_LDA}
\end{eqnarray}
Equations (\ref{eq:spins:kuebler1}) with $v_0({\bf r})$ and
$\Delta v({\bf r})$ given by Eq. (\ref{eq:spins:v_0_LDA}) 
are the KS equations which result from the approach by K\"ubler and
co-workers\cite{Kuebler,Sticht}. Here they are written in a form appropriate for the QHE. 
We now make
the extension to an ensemble DFT by introducing
occupation numbers $f_{i,\alpha}$ for the two states $\alpha$ 
in each spin-doublet $i$, and by
taking
\begin{equation}
\rho_{\sigma,\sigma'}({\bf r})=\sum_{i,\alpha}f_{i,\alpha}\psi_{i,\alpha}^\dagger({\bf r})
\hat S^\dagger_\sigma\hat S_{\sigma'}\psi_{i,\alpha}({\bf r}),
\label{eq:spins:ensemble_rho}
\end{equation}
where $\hat S_\sigma$ projects out the $\sigma$-component of the spinor
$\psi_{i,\alpha}({\bf r})$. IN ordinary DFT the occupancies $f_{i,\alpha}$
are zero or one. In our ensemble DFT calculations, we obtain fractional occupation
numbers using a method of running averages described in our 
earlier work\cite{Heinonen1,Heinonen2}.

\section{Formal results and numerical approximations}
\label{technical}
The usefulness of the LDA equations (\ref{eq:spins:kuebler1}), (\ref{eq:spins:v_0_LDA}),
and
(\ref{eq:spins:ensemble_rho}) ultimately depends on the
availability of good approximations for $\epsilon_{\rm xc}(\nu,\xi)$, 
the exchange-correlation energy
per particle of a homogeneous system of filling factor $\nu$ and polarization
$\xi$. We have previously\cite{Lubin} described our first attempt at constructing
an approximate energy surface for the QHE. 
In this section, we will describe in some detail
our work to improve on that approximation. In particular, we have
extended the exchange-correlation energy to the range $\nu>1$ and also better
incorporated electron-hole symmetry.

We start, as in our earlier work\cite{Lubin}, 
by approximating $\epsilon_{\rm xc}(\nu,\xi)$ as
\begin{equation}
\epsilon_{\rm xc}(\nu,\xi)=\epsilon^{\rm s}_{\rm xc}(\nu)+
\delta\epsilon_{\rm xc}(\nu) f(\xi)+\epsilon^{\rm C}(\nu).
\label{eq:technical:exc1}
\end{equation}
Here, $\epsilon^{\rm s}_{\rm xc}(\nu)$ is a smooth interpolation formula
for the ground state energy of polarized QHE systems, $\epsilon^{\rm C}(\nu)$
gives the cusps at the FQHE fractions, and $f(\xi)$ is
an interpolation formula obtained from considering only exchange in two
dimensions,
\begin{equation}
f(\xi)={(1+\xi)^{3/2}+(1-\xi)^{3/2}-2\sqrt{2}\over 2-2\sqrt{2}},
\label{eq:technical:f_xi}
\end{equation}
and $\delta\epsilon_{\rm xc}(\nu)$ is the difference in energy between
the fully polarized and the completely unpolarized ($\xi=0$) system at
filling factor $\nu$. Some values for this latter quantity can be obtained
from the literature\cite{Zhang,Maksym,Chakraborty}, 
and the value for $\delta\epsilon_{\rm xc}(\nu=1)$ will be 
fixed later in this section using the spin stiffness. We then use a 
spline fit to tie all these values of $\delta_{\rm xc}(\nu)$ together
to a continuous function. We found earlier\cite{Lubin} that the interpolation formula
for $\epsilon_{\rm xc}^{\rm s}(\nu)$ 
given by Fano and Ortolani\cite{Fano} 
together with our approximation for $\epsilon^{\rm C}(\nu)$
gave excellent 
agreement between our
DFT approach and numerical diagonalizations\cite{Yang} for small spin polarized
systems. However, this
interpolation formula was given only for $\nu\leq1$. 
To extend it to $\nu>1$, we have performed extensive
numerical diagonalizations for toroidal spin-polarized 
systems of eight, nine, ten, and eleven particles
in the two lowest Landau levels\cite{Yoshioka}. The data obtained from these
calculations reveal a cusp in the exchange-correlation energy at $\nu=1$.
This cusp is due to the fact that the exchange-correlation energy per particle 
added to the second Landau level, after the lowest Landau level has 
filled up, is different from the exchange-correlation
energy per particle for the filled lowest Landau level. We have confirmed this with
analytical calculations (below), 
and constructed a simple analytical model to fit
the numerical data. Figure \ref{fig:num_energies}
depicts the data from the numerical diagonalizations and the analytical fit.

We now briefly present the analytical calculation of 
the cusp in the exchange-correlation energy at
$\nu=1$ for spin-polarized systems.  
We will do the calculation in a truncated Hilbert space, and take the
state at $\nu=1$ to be a Slater determinant 
of the lowest Landau level
single-particle wavefunctions. This is the exact ground state of the
system restricted to the lowest Landau level, and the interaction
energy is $E_0=N_{\rm el}\epsilon_{\rm xc}(\nu=1,\xi=1)=-N_{\rm el}\sqrt{\pi\over8}
{e^2\over\epsilon_0 l_B}$, where $N_{\rm el}$ 
is the number
of electrons. For $N_{\rm el}+1$ particle we
consider only eigenstates with $N_{\rm el}$ particles occupying a filled lowest
Landau level plus one particle in the $n=1$ Landau level. The lowest-energy
state consisting of $N_{\rm el}+1$ particles
is then a linear combination of these degenerate Slater determinants with 
a uniform density. The direct (Hartree) energy of this state is canceled
by a uniform positive background charge density. 
We can calculate the exact energy of the lowest-lying state (exact in the reduced
Hilbert space used here) by considering the exchange 
interaction between a single particle
in Landau level $n=1$ at momentum $k_0=0$ and a particle of 
momentum $k$ in the lowest Landau level $n=0$. This
energy is (using the Landau gauge)
\begin{eqnarray}
\epsilon_{1}(k) & = & -\int\int d^2r\,d^2r'\,{1\over L_y^2\pi\l_B^2}
e^{-\frac{1}{2l_B^2}(x-x_k)^2}e^{-iky}
e^{-\frac{1}{2l_B^2}{x'}^2}
{H_1(x'/l_B)\over \sqrt{2}}V({\bf r}-{\bf r}')\nonumber\\
&&\times{H_1(x/l_B)
\over\sqrt{2}}
e^{-\frac{1}{2l_B^2}x^2}e^{iky'}e^{-\frac{1}{2l_B^2}(x'-x_k)^2}.
\label{eq:technical:e_x1}
\end{eqnarray}
Here, $L_y$ is the length of the system (taken to infinity at the
end of the calculation), $H_n$ is the $n^{th}$ Hermite polynomial,
$V(r)=e^2/(\epsilon_0|{\bf r}-{\bf r}'|)$ the Coulomb interaction,
and $x_k=l_B^2k$ is the centerpoint of the harmonic oscillator of momentum $k$.
The integrals in Eq. (\ref{eq:technical:e_x1}) can be
evaluated to give the result
\begin{equation}
\epsilon_1(k)=-{e^2\over2L_y\epsilon_0}e^{-k^2l_B^2/4}{k^2 l_B^2\over4}
\left[K_0(k^2l_B^2/4)+K_1(k^2l_B^2/4)
\right],
\label{eq:technical:e_x2}
\end{equation}
where $K_n$ is the modified Bessel function of order $n$. We then
finally integrate over all states $k$ in the lowest Landau level and obtain
\begin{equation}
\epsilon_{1}=-\frac{1}{2}\sqrt{\pi\over8}{e^2\over\epsilon l_B}.
\label{eq:technical:e_x}
\end{equation}
The cusp in ground state energy at $\nu=1$ comes from the fact
that the exchange energy of adding one particle to a system with a 
full lowest Landau level, $\epsilon_1$, is different from the exchange  energy per particle
in the lowest Landau level. The cusp gives rise to a discontinuity
in the chemical potential at $\nu=1$, which we need to evaluate.
The chemical potential at $\nu=1^-$ is
$-\sqrt{\pi/2}e^2/(\epsilon_0 l_B)$, and the chemical potential at $\nu=1^+$ is
\begin{eqnarray}
\mu(\nu=1^+) & = & \left.{\partial E\over\partial N_{\rm el}}\right|_{\nu=1^+}
=E(N_{\rm el}+1)-E(N_{\rm el})\nonumber\\
& = & \left[N_{\rm el} \epsilon_{\rm xc}(\nu=1,\xi=1)+\epsilon_1\right]-N_{\rm el}
\epsilon_{\rm xc}(\nu=1,\xi=1)
 = \epsilon_1=-\frac{1}{2}
\sqrt{\pi\over8}{e^2\over\epsilon_0l_B}.
\end{eqnarray}
Therefore, the discontinuity in chemical potential at $\nu=1,\xi=1$ is
\begin{equation}
\Delta\mu=\frac{3}{2}\sqrt{\pi\over8}{e^2\over\epsilon_0l_B}.
\end{equation}
Since the exchange-correlation potential for a polarized system is
$V_{\rm xc}(\nu)=\partial\left[\nu\epsilon_{\rm xc}(\nu,\xi=1)\right]/(\partial
\nu)=\mu(\nu)$ this discontinuity also appears in the 
exchange-correlation potential at $\nu=1$.

In order to construct an exchange-correlation energy surface
$\epsilon_{\rm xc}(\nu,\xi)$ which gives a workable
approximation for $0\leq\nu\leq2$ and $0\leq\xi\leq1$ we
first construct an analytic approximation for the exchange-correlation
energy at $\nu>1$ and $\xi=1$. We use a simple model in which we
write
\begin{equation}
\epsilon_{\rm xc}(\nu>1,\xi=1)=a{\nu-1\over\nu}\epsilon_{\rm xc}(\nu-1,\xi=1)
+\frac{c}{\nu}+b.
\label{eq:technical:an_appr}
\end{equation}
This model is motivated by the fact that, as a first approximation, the interaction 
energy 
of a system with a full lowest Landau level and
$N_1$ particles in the $n=1$ Landau level is approximately
equal to the interaction energy of the full lowest Landau level plus
the interaction energy of $N_1$ electrons in the lowest Landau level. The
constants $b$ and $c$ then fixes the slope and value
of the exchange-correlation energy at $\nu=1^+$ and $\xi=1$, and $a$ is used
to adjust this model to the numerical data. Fixing slope and value gives
\begin{equation}
b=\frac{3}{2}\epsilon_{\rm xc}(\nu=1,\xi=1),
\end{equation}
and
\begin{equation}
c=-\frac{1}{2}\epsilon_{\rm xc}(\nu=1,\xi=1),
\end{equation}
respectively. A good fit to the numerical data is given by $a=2$.

Finally, we consider the exhange-correlation energy at $\xi=0$. When the
system is restricted to the lowest Landau level, particle-hole symmetry yields
\begin{equation}
\nu\epsilon_{\rm xc}(\nu,\xi)-\nu\epsilon_{\rm xc}(\nu=1,\xi=1)
=(2-\nu)\epsilon_{\rm xc}\left[2-\nu,(\nu_\uparrow-\nu_\downarrow)/(2-\nu)\right]-(2-\nu)
\epsilon_{\rm xc}(\nu=1,\xi=1),
\label{eq:technical:e-h_symmetry}
\end{equation}
where
\begin{equation}
\nu=\nu_\uparrow+\nu_\downarrow>1
\end{equation}
and
\begin{equation}
\xi={\nu_\uparrow-\nu_\downarrow\over\nu_\uparrow+\nu_\downarrow}.
\end{equation}
When the restriction of the system to the lowest Landau level is lifted, 
this symmetry is no longer exact
because of inter-Landau level quantum fluctuations. 
However, we assume that it is only slightly violated, and
construct $\delta\epsilon_{\rm xc}(\nu)$ so that 
$\epsilon_{\rm xc}(\nu,\xi)$ respects this symmetry at $\xi=0$.
Using the form Eq. (\ref{eq:technical:exc1}) and Eq. (\ref{eq:technical:e-h_symmetry})
at $\xi=0$ and $\nu>1$ then yields
\begin{eqnarray}
\delta\epsilon_{\rm xc}(\nu) & = & {\left[2-\nu\right]\left[
\epsilon_{\rm xc}^{\rm s}(2-\nu)+\epsilon^{\rm C}(2-\nu)+
\delta\epsilon_{\rm xc}(2-\nu)\right]-2(1-\nu)\epsilon_{\rm xc}(\nu=1,\xi=1)
\over
\nu}\nonumber \\
&&-\epsilon_{\rm xc}^{\rm s}(\nu)-\epsilon^{\rm C}(\nu)
\label{eq:technical:e-h_symm2}
\end{eqnarray}
for $\nu>1$.
Furthermore, at $\xi=0$ the exchange-correlation energy per particle
has a continuous derivative at $\nu=1$, which gives
\begin{equation}
\left.{d\left(\delta\epsilon_{\rm xc}(\nu)\right)\over d\nu}\right|_{\nu=1}
=-\epsilon_{\rm xc}(\nu=1,\xi=1)-\delta\epsilon_{\rm xc}(\nu=1).
\label{eq:technical:e-h_symm_deriv}
\end{equation}
Equations (\ref{eq:technical:exc1}), (\ref{eq:technical:f_xi}), 
(\ref{eq:technical:an_appr}), (\ref{eq:technical:e-h_symm2}), and
(\ref{eq:technical:e-h_symm_deriv}), together with the data points for 
$\delta\epsilon_{\rm xc}(\nu)$ for $\nu\leq1$ then define our exchange-correlation
energy surface. A F90 subroutine package which evaluates the exchange-correlation
energy and the exchange-correlation potentials for 
given $(\nu_\uparrow,\nu_\downarrow)$ is available from the authors. In this package,
used in our calculations, we dropped the terms $\epsilon^{\rm C}$ from 
Eq. (\ref{eq:technical:e-h_symm2}). This is of no consequence in our calculations
presented here in which the total filling factor was never much greater than unity.

To conclude this section, we show that the LDA does
indeed give a spin stiffness; requiring this stiffness to be correct helps
constrain $\delta\epsilon_{\rm xc}(\nu=1)$. We start by considering
the total exchange-correlation energy of the system in the LDA,
\begin{equation}
E_{\rm xc}=\int\,n({\bf r})\epsilon_{\rm xc}\left[\nu({\bf r}),\xi({\bf r})\right]
\,d^2r,
\label{eq:technical:E_xc}
\end{equation}
where $\nu({\bf r})$ and $\xi({\bf r})$ are obtained from the local
eigenvalues $n_\uparrow({\bf r})$ and $n_\downarrow({\bf r})$ 
of the single-particle density matrix\cite{Sticht}. We calculate in the LDA 
the change in
exchange-correlation energy of an initially infinite, homogeneous, 
fully polarized system at
$\nu=1$ in response to a gentle spin twist. The applied spin twist changes the
local eigenvalues $n_\sigma({\bf r})$ of the single-particle
density matrix, and thus $\nu({\bf r})$ and
$\xi({\bf r})$.  We write
\begin{equation}
\nu({\bf r})=1+\delta\nu({\bf r}),\quad
\xi({\bf r})=1+\delta\xi({\bf r}).
\end{equation}
The change in exhange-correlation energy is then 
(with $n({\bf r})=\nu({\bf r})/(2\pi l_B^2)$)
\begin{equation}
\Delta E_{\rm xc}=\int\left[\frac{1}{2\pi}+\delta n({\bf r})\right]
\epsilon_{\rm xc}\left[1+\delta\nu({\bf r}),1+\delta\xi({\bf r})\right]
\,d^2r-\frac{1}{2\pi}
\int\epsilon_{\rm xc}(1,1)\,d^2r.
\label{eq:technical:Delta_Exc}
\end{equation}
From the work
of Moon {\em et al.,}\cite{Moon} we know that the density variation is of second order
in the gradient of the spin-rotation angle. Furthermore, general symmetry
considerations give
\begin{equation}
\delta\xi({\bf r})\propto\left[\nabla{\bf \Omega}({\bf r})\right]^2,
\label{eq:technical:delta_xi}
\end{equation}
where ${\bf \Omega}({\bf r})=\hat z\times{\bf m}({\bf r})$ is the angle through
which the spin density is rotated, and ${\bf m}({\bf r})$ is a unit vector
parallel to the local spin density. We then expand $\Delta E_{\rm xc}$ in powers
of $|\nabla{\bf \Omega}|$, and obtain to second order in $|\nabla{\bf \Omega}|$
\begin{equation}
\Delta E_{\rm xc}=\int\,\delta n({\bf r})\epsilon_{\rm xc}(1,1)\,d^2r
+\frac{1}{2\pi}\int\,\delta\nu({\bf r})
\left.{\partial\epsilon_{\rm xc}\over\partial\nu}\right|_{\nu=1,\xi=1}\,d^2r
+\frac{1}{2\pi}\int\,\delta\xi({\bf r})
\left.{\partial\epsilon_{\rm xc}\over\partial\xi}\right|_{\nu=1,\xi=1}\,d^2r.
\label{eq:technical:Delta_Exc2}
\end{equation}
The point here is that an LDA in $\nu({\bf r})$ and $\xi({\bf r})$ contains
spatially varying polarization $\xi({\bf r})$, and this corresponds to
gradients in the magnetization density according to Eq. (\ref{eq:technical:delta_xi}).
The first two terms on the left-hand side describe a change in $E_{\rm xc}$
due a change in the density. For the spin twist we are considering
$\delta n({\bf r})$ (and $\delta\nu({\bf r})$) integrates to zero since
no net charge is added to the system. The last term in 
Eq. (\ref{eq:technical:Delta_Exc2}) describes the change in $E_{\rm xc}$ due
to a change in the polarization. This term gives the spin stiffness. In general,
the spin stiffness $\rho_s$ is defined by the leading term in an expansion
of the energy in gradients of the magnetization angle $\bf\Omega$:
\begin{equation}
E_s=\frac{1}{2}\rho_s\int \left[\nabla{\bf \Omega}({\bf r})\right]^2\,d^2r.
\label{eq:technical:rho_s}
\end{equation}
We obtain from
Eqs. (\ref{eq:technical:delta_xi}) and (\ref{eq:technical:Delta_Exc2})
\begin{equation}
\Delta E_{\rm xc}\propto\frac{1}{2\pi}
\left.{\partial\epsilon_{\rm xc}\over\partial\xi}\right|_{\nu=1,\xi=1}
\int\,\left[\nabla{\bf \Omega}({\bf r})\right]^2\,d^2r.
\label{eq:technical:Delta_Exc3}
\end{equation}
By comparing Eqs. (\ref{eq:technical:rho_s}) and (\ref{eq:technical:Delta_Exc3})
we see that $\rho_s\propto(\partial\epsilon_{\rm xc}/\partial\xi)$, and it
remains to work out the constant of proportionality. To this end, we consider
a system of spin-polarized electrons confined to the lowest Landau level. We use
the Landau gauge ${\bf A}({\bf r})=(0,Bx,0)$ in which the single-particle
basis functions are
\begin{equation}
\psi_k(x,y)=e^{iky}\phi_k(x)=
{e^{iky}e^{-\frac{1}{2l_B^2}(x-x_k)^2}\over
{\sqrt{L_y}}{\sqrt{\sqrt\pi l_B}}},
\label{eq:technical:basis}
\end{equation}
where
\begin{equation}
x_k=l_B^2k,\quad k=2\pi n/L_y,\quad n=0,\pm1,\pm2,\ldots.
\end{equation}
In the initial state, all single-particle
spinors are
\begin{equation}
\left(\begin{array}{c}
\psi_k(x,y)\\
0
\end{array}\right).
\end{equation}
We take the spinors of the spin-rotated state to be
\begin{equation}
\left(
\begin{array}{c}
\cos\frac{\theta_k}{2}\psi_k(x,y)\\
\sin\frac{\theta_k}{2}\psi_{k-\Delta k}(x,y)
\end{array}
\right),
\end{equation}
where $\Delta k=2\pi\Delta n/L_y$, with $\Delta n$ a fixed integer.
This state is the rectangular analogue of a rotationally symmetric skyrmion.
In order to calculate the exchange-correlation energy of this state within the
LDA, we then need to find the eigenvalues of the single-particle
density matrix
\begin{equation}
\rho_{\sigma,\sigma'}({\bf r})=\left(
\begin{array}{cc}
\sum_k\cos^2\frac{\theta_k}{2}|\psi_k(x,y)|^2 & \sum_k\sin\frac{\theta_k}{2}
\cos\frac{\theta_k}{2}\psi_k(x,y)\psi^*_{k-\Delta k}(x,y) \\
\sum_k\sin\frac{\theta_k}{2}\cos\frac{\theta_k}{2}
\psi^*_k(x,y)\psi_{k-\Delta k}(x,y) &
\sum_k\sin^2\frac{\theta_k}{2}|\psi_{k-\Delta k}(x,y)|^2
\end{array}
\right).
\end{equation}
The eigenvalues are then readily obtained as
\begin{eqnarray}
n_{\uparrow,\downarrow}(x) & = &
\left[\sum_k\cos^2\frac{\theta_k}{2}\phi_k^2(x)+\sum_k\sin^2\frac{\theta_k}{2}
\phi_{k-\Delta k}\right]\nonumber \\
&& \pm\frac{1}{2}\left\{
\left[\sum_k\cos^2\frac{\theta_k}{2}\phi_k^2(x)-\sum_k\sin^2\frac{\theta_k}{2}
\phi_{k-\Delta k}\right]^2
+\left[\sum_k\sin\theta_k\cos\theta_k
\phi_k(x)\phi_{k-\Delta k}(x)\right]^2\right\}^{1/2}.
\label{eq:technical:eigenvalues}
\end{eqnarray}
To continue, we will make use of the following results
\begin{eqnarray}
\sum_k\phi_k^2(x)&=&{1\over2\pi l_B^2},\nonumber\\
\sum_k\left(x-x_k\right)^2\phi_k^2(x) & = 
& {1\over2\pi l_B^2}{l_B^2\over2}
={1\over4\pi}.
\label{eq:technical:results}
\end{eqnarray}
and also the Taylor expansion
\begin{equation}
\phi_k(x)\phi_{k-\Delta k}(x)\approx\phi_k(x)\left[
\phi_k(x)-\Delta k{\partial\phi_k(x)\over\partial k}+
\frac{1}{2}(\Delta k)^2{\partial^2\phi_k(x)
\over\partial k^2}\right].
\label{eq:technical:taylor}
\end{equation}
The second term on the right-hand side of Eq. (\ref{eq:technical:taylor}) then
vanishes when integrated over $k$. The third term contains 
\begin{equation}
(\Delta k)^2=\left({2\pi\over L_y}\right)^2\left(\Delta n\right)^2,
\end{equation}
which vanishes in the thermodynamic limit for fixed $\Delta n\not=0$. Therefore,
we can approximate
\begin{equation}
\sum_k\phi_k(x)\phi_{k-\Delta k}(x)
\approx\sum_k\phi_k^2(x)={1\over2\pi l_B^2}.
\end{equation}
Next, we assume that $\theta_k$ is slowly varying as a function of $k$. This means
that in expressions like
\begin{equation}
\sum_k\cos^2\frac{\theta_k}{2}\phi_k^2(x)={L_y\over2\pi}\int\,dk
\frac{\cos^2\theta(k)}{2}\phi_k(x)
\end{equation}
where $\phi_k(x)$ as a function of $k$ is sharply peaked about $k=x/l_B^2$, we
can expand to the trigonometric function as a function of $k$ to second order 
in $k$ about $k=x/l_B^2$.
When integrated over $k$, all first-order terms containing $d\theta/dk$ then
vanish. Using Eq. (\ref{eq:technical:results}) and
$d^2\theta/dk^2=l_B^4d^2\theta/(dx^2)$ we obtain, after a little
algebra,
\begin{equation}
n_{\uparrow,\downarrow}=\frac{1}{2}\frac{1}{2\pi l_B^2}
\pm\frac{1}{2}\frac{1}{2\pi l_B^2}\left[1-\frac{l_B^2}{4}\left({d\theta\over dx}
\right)^2\right\}.
\label{eq:technical:n_values}
\end{equation}
From the definition of the polarization $\xi({\bf r})$ we then finally have
\begin{equation}
\delta\xi({\bf r})=-{2n_\downarrow({\bf r})\over n_\uparrow({\bf r})
-n_\downarrow({\bf r})}=-{l_B^2\over4}\left({d\theta\over dx}\right)^2.
\label{eq:technical:delta_xi2}
\end{equation}
Inserting this into Eq. ({\ref{eq:technical:Delta_Exc2}) yields
\begin{equation}
\Delta E_{\rm xc}=-{1\over 4\pi}\left.{\partial\epsilon_{\rm xc}\over\partial\xi}
\right|_{\nu=1,\xi=1}{1\over2}\int\left({d\theta\over dx}\right)^2\,d^2r.
\label{eq:technical:Delta_Exc4}
\end{equation}
By comparing Eq. (\ref{eq:technical:rho_s}) and Eq. (\ref{eq:technical:Delta_Exc4})
we then obtain for the spin stiffness
\begin{equation}
\rho_s=-{1\over4\pi}\left.{\partial\epsilon_{\rm xc}\over\partial\xi}\right|_{\nu=1,\xi=1}
.
\label{eq:technical:rho_s2}
\end{equation}
In fact, the calculation is readily generalized to an arbitrary spin-polarized
filling factor $\nu_0$, so in general we have
\begin{equation}
\rho_s=-{\nu_0\over4\pi}\left.{\partial\epsilon_{\rm xc}(\nu,\xi)\over\partial\xi}
\right|_{\nu=\nu_0,\xi=1}.
\label{eq:technical:rho_s_general}
\end{equation}
In our approximation for the exchange-correlation energy, we can then fix
$\delta\epsilon_{\rm xc}(\nu=1)$ (for which there is no known value)
by requiring that the LDA spin stiffness
Eq. (\ref{eq:technical:rho_s2}) equals the known value\cite{Moon} for the spin stiffness
at $\nu=1$, 
\begin{equation}
\rho_s={1\over16}{1\over\sqrt{2\pi}}{e^2\over\epsilon_0l_B}.
\end{equation}
This yields
\begin{equation}
\delta\epsilon_{\rm xc}(\nu=1)=\sqrt{\pi\over8}{2-\sqrt{2}\over3}
{e^2\over\epsilon_0l_B}\approx0.1224{e^2\over\epsilon_0l_B}.
\label{eq:technical:delta_e_value}
\end{equation}
As an indicator how good this value is for $\delta\epsilon_{\rm xc}(\nu=1)$, we
consider the state at $\nu=1,\xi=0$ in an approximation in which the
spin-up and spin-down particles are completely uncorrelated. In this case
the energy per particle of that state is the same as the ground state energy
per particle at $\nu=\frac{1}{2},\xi=1$, which is about\cite{Fano}
$-0.469$ $e^2/(\epsilon_0l_B)$.
Neglecting the correlations between spin-up and spin-down particles
should lead to an overestimate of the energy per particle.
This approximation yields a difference in energy per particle between $\nu=1,\xi=0$ and
$\nu=1,\xi=1$ of $\delta\epsilon_{\rm xc}(\nu=1)=0.1577$ $e^2/(\epsilon_0l_B)$. 
In view of the fact that this is most likely an overestimate of
the energy per particle at $\nu=1,\xi=0$, we can
conclude that the value $\delta\epsilon_{\rm xc}(\nu=1)=0.1224e^2/(\epsilon_0 l_B)$ 
obtained
from the spin stiffness is very reasonable. The value 
$\delta\epsilon_{\rm xc}(\nu=1)=0.1224\,e^2/(\epsilon_0 l_B)$
is the one we used in our calculations.

\section{Edge reconstructions of QHE dots}
\label{dot}
We have applied the ensemble spin DFT approach to circularly symmetric
QHE dots and studied the edge reconstructions of such systems as a function
of edge stiffness. We use a model in which the confining potential is
supplied by a uniform positive background charge density. 
The edge is modeled by a `graded edge' in which the positive
background charge density goes to zero linearly as a function of radial
coordinate $r$ over a distance $w$. The total integrated positive charge
is fixed and equal to the total electron charge; this then determines the
radial distance at which the positive charge density starts to decrease.
Even though concerns have been raised that this particular confinement is
non-generic\cite{Karlhede2} it is the one which has been studied the
most, and we chose it as a model confinement for
comparisons with other work.

It is known from Hartree-Fock calculations\cite{Karlhede1} of $\nu=1$ Hall bars with a 
similar graded edge that for a sharp edge ($w$ small enough), 
the electron gas is completely polarized and its density falls to zero
abruptly near the edge. However, as the edge confinement softens, the
electron gas develops an instability to a spin textured edge for smaller
values of $\tilde g$ (the ratio of Zeeman to Coulomb energy), 
or a charge density wave\cite{Franco,Karlhede2} for larger values of $\tilde g$, 
with the density modulated
along the edge. Both instabilities break the translational invariance along
the edge: in the spin-textured edge the spin density is modulated along the
edge while the total charge density is constant; in the charge-density wave
edge the spin density is constant along the edge while the total charge
density is modulated. For Hall bars at bulk filling $\nu=1/3$, an effective
field calculation\cite{Karlhede1} (here, Hartree-Fock calculations are obviously not
applicable) also shows an instability to a spin textured edge as
the confinement is softened. 

We would expect the analogous instabilities
to occur for the circular dots. For stiff confinements, the electron
density forms a so-called maximum density droplet (MDD), in which the
electron gas is completely polarized with a filling factor which is unity in
the bulk and which rapidly falls to zero at the edge (the MDD
is the minimum angular momentum state for a spin-polarized system in the
lowest Landau level). As the confinement is softened, 
the edge should develop a spin
textured or charge density wave instability. Indeed, our previous spin
DFT calculations\cite{Lubin}, in which the magnetization direction was constrained to
be fixed parallel to the external magnetic field, revealed that the MDD
becomes unstable towards the formation of
a partially polarized edge as the confinement was softened. This gave
a variational bound (within the LDA) showing that the spin-polarized edge
is not the ground state when the confinement is soft. The phase boundaries of the
MDD that we obtained were in good qualitative agreement with Hartree-Fock calculations
for Hall bars\cite{Karlhede1}, 
although the obtained values of $\tilde g$ at which the polarization
was destroyed were much smaller than those from the Hartree-Fock calculation for
Hall bars. We 
speculated that this difference is due to the different geometries or edge confinements.
Our results were in rather good agreement with numerical diagonalizations
using parabolic confinement\cite{Yang}. 
The calculations presented here support the argument that the differences were due
to the different confinements used.

We have now extended our ensemble DFT calculations to include spin textured edges both
for the MDD and for $\nu=1/3$ droplets. 
Our new results show that when the edge
becomes partially polarized, a spin textured edge has lower energy than
one with constant direction of the spin density.
In the calculations presented here, we have only considered states which
do not break cylindrical symmetry of the charge density, and for which the
azimuthal angle of the spin direction changes at most by $2\pi$ along any simple
closed path. This excludes charge density wave instabilities and
spin textures with topological charge greater than $\pm e$, but
imposing these symmetries
simplifies the calculations a great deal. For example, the Hartree potential
is easy to calculate for circularly symmetric charge densities, but
considerably more tedious if that symmetry is broken. Nevertheless, the calculations
we have performed satisfy the two most important criteria we wanted to
establish: to demonstrate the usefulness of the spin
ensemble DFT approach to general QHE fillings, and 
to establish through a variational bound (within the LDA)
that for softer edges, the spin-polarized edge or an edge with constant
spin quantization axis has higher energy than a spin-textured system in which
the spin quantization axis is tumbling.

We start with the ensemble spin KS equations Eq.~(\ref{eq:spins:kuebler1})
with the potentials $v_0({\bf r})$ and $\Delta v({\bf r})$ given by Eqs.~
(\ref{eq:spins:v_0_LDA}), and the single-particle density matrix found self-consistently
using Eq.~(\ref{eq:spins:ensemble_rho}).
We now make the simplifying assumption that the polar angle $\theta({\bf r})$ of the
spin density is a function of the radial coordinate alone, $\theta({\bf r})=\theta(r)$,
and that the azimuthal angle $\varphi({\bf r})$
is of the form
\begin{equation}
\varphi({\bf r})=v\phi({\bf r}),
\label{eq:dots:spin_azimuthal}
\end{equation}
where $v$ is an integer and $\phi({\bf r})$ is the azimuthal angle of ${\bf r}$
in a planar polar coordinate system. In other words, we will restrict the
modulation of the spin density along the edge to have a single Fourier
component along the edge. With Eq.~(\ref{eq:dots:spin_azimuthal}) inserted
into the expressions Eq.~(\ref{eq:spins:kuebler1}) for 
$v_0({\bf r})$ and $\Delta v({\bf r})$
the spin-diagonal coupling conserves orbital angular
momentum, while the spin off-diagonal coupling couples up-spin
states $\psi_{m+v,\uparrow}$ with angular momentum $\hbar(m+v)$ 
to down-spin states $\psi_{m,\downarrow}$ with angular momentum $\hbar m$:
\begin{equation}
\int \cos\theta(r)\psi_{m,\downarrow}^*({\bf r)}\Delta v(r)e^{-iv\phi}
\psi_{m+v,\uparrow}({\bf r})\,d^2r.
\label{eq:dots:spin_coupling}
\end{equation}
This coupling is of the same form as that in the Hartree-Fock equations
studied first by Fertig {\em et al}\cite{Fertig}. However, in the Hartree-Fock equations
the off diagonal coupling was provided by an exchange integral, while our 
Eq.~(\ref{eq:dots:spin_coupling}) also includes correlation effects within the LDA.
Since the spin-diagonal coupling
conserves angular momentum, the diagonal elements of the single-particle
density matrix are circularly symmetric so that the total density is
circularly symmetric. We would like to point out that lifting the restriction
Eq.~(\ref{eq:dots:spin_azimuthal}) allows for the possibility of breaking the
circular symmetry of the charge density, which allows for charge-density waves in
addition to spin-charge structures.

We have
solved the KS equations by expanding the spatial parts of each spinor
$\psi_{i,\alpha}({\bf r})$ in the single-particle angular momentum basis functions 
\begin{equation}
\psi_{m,n}({\bf r})={1\over\sqrt{2\pi}l_B}\sqrt{n!\over(n+m)!}
L_m^n\left(\frac{r^2}{2l_B}\right)
e^{im\phi},
\label{eq:dots:basis_functions}
\end{equation}
for
the cylindrical gauge ${\bf A}({\bf r})=\frac{1}{2}B(x\hat {\bf y}-y\hat {\bf x})$ with
$L_m^n$ the associated Laguerre polynomials. We kept
up to the $n=4$ Landau level in our calculations, and up to 120
angular momentum states (up to 1,200 single-particle states). For the $\nu=1$ system we
performed calculations of 40 and 70 particles in a magnetic field of 3.5 T with
the bare Land\'e factor $g^*$ varying from 0.1 to 1.5 ($\tilde g$ then varied from
about 0.002 to 0.036, encompassing experimentally accessible values). The reason
to keep the magnetic field relatively low, but still at an experimentally
realistic value, was to fully include the effects of Landau level mixing
typical in experimental system. We
performed the calculations for $v=0,\pm1$. The results
can be summarized as follows: For small values of $w$, {\em i.e.\/} stiff confinement,
the edge is spin polarized. As $w$ increases, the $v=0$ channel becomes partially
polarized. However, at the same value of $w$,
the $v=1$ channel attains a lower energy with a nontrivial
spin-charge texture. For the system sizes studied here, the value of $w$ at which
the instability occurred is $w\approx7l_B$. This is in quite good
agreement with the Hartree-Fock calculations of 
Karlhede {\em et al.}\cite{Karlhede1}. For
a semi-infinite Hall bar, they found the onset to a charge-spin textured edge
occurring at about $w\approx7\,l_B$ -- $8\,l_B$ in the range of $\tilde g$ from
zero to about 0.03. As illustrations, figures \ref{fig:v0_nu_1_dens} and 
\ref{fig:v1_nu_1_dens} depict charge densities and spin rotation angle for
$w=8\,l_B$ and $g=1.00$. The $v=0$ channel (Fig.~\ref{fig:v0_nu_1_dens}) has a
small minority-spin density near the edge of the system, in qualitative
agreement with our earlier calculations (the bump in density at the
edge is characterisic of all confined QHE systems for which the confining potential
is not macroscopically smooth). However, the $v=1$ channel (Fig. \ref{fig:v1_nu_1_dens}), 
which has lower energy,
has a locally polarized spin density everywhere and a non-trivial spin texture, with the
spin rotation angle $\theta(r)$ rising from 0 to about $\pi/2$ at the edge of the
system. Note that for a bounded system there is no topological constraint on the
spin rotation angle, as there is in an infinite system where the Pontryagin index 
has to be an integer.

We have also studied a FQHE droplet at bulk filling
of 1/3. While we are quite confident about our exchange-correlation energy
for spin-polarized FQHE systems, it is not clear to us how good it is for
arbitrary polarizations in the FQHE regime. For example, if we use the
published value obtained from four-electron numerical
diagonalizations\cite{Zhang} for $\delta\epsilon_{\rm xc}(\nu=1/3)$ of 
0.0017 $e^2/(\epsilon_0l_B)$,
the spin-stiffness in our model is $1.155\times10^{-4}\,e^2/(\epsilon_0l_B)$,
compared to the value of $9.23\times10^{-4}\,e^2/(\epsilon_0l_B)$
obtained from hypernetted-chain calculations\cite{Moon}. It is not clear
if the discrepancy between these two is mostly due to finite-size effects
in the numerical diagonalizations, or our model Eqs.~(\ref{eq:technical:exc1}) and
(\ref{eq:technical:f_xi}) of the exchange-correlation energy. We have
also performed numerical diagonalizations of six electrons at $\nu=1/3$. For
six electrons, we obtain $\delta\epsilon_{\rm xc}(\nu=1/3)=0.00465$ $e^2/(\epsilon_0 l_B)$ -- 
an increase of almost a factor of three from the four-electron result. This
clearly shows that the energy per particle at total spin zero is much more
sensitive to system size than the energy of the polarized ground state. Encouraged
by our numerical results, we then we fixed
$\delta\epsilon_{\rm xc}(\nu=1/3)$ to $0.0136\,e^2/(\epsilon_0l_B)$  in order to
have a simple model which gives a spin stiffness in agreement with 
the hypernetted chain calculations.
We performed the calculation with 40 particles and the four lowest Landau levels,
and up to 170 single-particle angular momentum states, both for the choice of confinement
discussed earlier and parabolic confinement.

Our calculations indicate that for a confinement provided
by the positive background charge density, the system has an instability from a spin-polarized
edge to a spin-textured edge at an edge width of $w\approx4\,l_B$. Again, this
compares rather well with the results of Karlhede {\em et al.}\cite{Karlhede1} Using
effective-field theories, they found an instability to a spin-textured edge
at about $w=3.0\,l_B$ for $\tilde g=0.04$. Note that the effective-field
theory tends to underestimate the value of $w$ for which there is an
onset to spin-textured edges\cite{Karlhede1}. We also want to emphasize that in
contrast to the effective-field (and Hartree-Fock) theory, our ensemble spin
DFT is applicable to general inhomogeneous QHE system and includes the
effects of Landau level mixings.
As an example of our results, we show in figures \ref{fig:v0_nu_1/3_dens}
and \ref{fig:v1_nu_1/3_dens} charge densities and spin rotation angles for
a system with an edge width of $w=4$ and Land\'e factor of $g^*=1.00$ in a 
magnetic field of
12 T ($\tilde g\approx0.04$). The bump in total filling factor in the bulk of the
$v=0$ channel (Fig. \ref{fig:v0_nu_1/3_dens}) occurs quite generically
for FQHE droplets, both for the choice of confinement discussed here and for
parabolic confinement. In the case here, the system tries to take advantage of correlation
energy at $\nu=1/3$ and $\nu=2/5$ to as large extent as possible as the edge is made wider.
It does so by making the edge of the electron density sharper than that of the
background charge, and making regions of $\nu=1/3$ larger than what is needed to
accomodate all electron charge. The residual electron charge is piled up in a bump
reaching $\nu=2/5$. The $v=1$ channel, on the other hand, is locally completely
polarized with a spin rotation angle similar to the $\nu=1$ and $w=8$ system for $v=1$.

We have also performed calculations for parabolically confined systems in the FQHE regime.
The results are in this case more difficult to interpret with more complicated
structures in the electron density and spin textures. For example, we have found
that for some values of the
magnetic field
the $v=0$ channel can be a $\nu=1/3$ droplet with a large bump in density at the edge, while
the $v=1$ channel develops a hole with reversed spin density at
the center and has lower energy for a small range of magnetic field.

\section{Conclusions and Summary}
\label{end}
We have here presented an ensemble spin DFT approach to general inhomogeneous QHE systems.
This approach includes the spin degree of freedom, including non-collinear
spins, as well as Landau level mixing. On a formal level, we have
demonstrated that our ensemble spin DFT in the LDA can give the correct
spin stiffness at $\nu=1$. We have performed model calculations for 
circularly symmetric QHE dots in the integer and FQHE regime. These calculations show,
in agreement with Hartree-Fock and 
effective-field calculations\cite{Karlhede1,Franco,Karlhede2}, that the polarized system
develops an instability as the confinement is softened and that the spin-textured edge
attains a lower energy than the spin-polarized one. We have not included in our
calculations the possibilities of charge-density waves, 
which may occur\cite{Franco,Karlhede2}
instead of the spin textured edges at larger values of
$\tilde g$. 
Preliminary calculations of
parabolic dots show a surprisingly rich structure in spin- and charge-densities. This
indicates that quantum dots in the FQHE regime is a rich subject yet to be fully
explored.

We have spent a great deal of effort on improving our approximation for the exchange-correlation
energy $\epsilon_{\rm xc}(\nu,\xi)$. At the present, we are confident that we have a
very good approximation for spin polarized systems, and a good approximation for
aribitrary polarizations and $\nu\approx1$. We are less confident about
our exchange-correlation energy for arbitrary polarizations in the FQHE regime. Work
need still to be done to refine the exchange-correlation energy
for general FQHE systems. However, we are confident that the approach itself is robust
and accurate provided good approximations for the exchange-correlation energy exist.

\section{Acknowledgements}
The authors are greatly indebted to Professor E.K.U. Gross for pointing out the
work on non-collinear spins by K\"ubler and co-workers. We would also like
to thank Professor Walter Kohn for stimulating and challenging comments on our
ensemble DFT approach, and Professor A.H. MacDonald for being a general source 
of wisdom of the QHE to whom we can always turn for good comments and inspiration.
O.H. and M.D.J. are grateful for support from the NSF through grant DMR96-32141, and
O.H. would also like to thank S. \"Ostlund and M. Jonson at
Chalmers University of Technology and G\"oteborg University
for support and hospitality during a sabbatical leave where this work was started, and
P. Hawrylak and C. Dharma-Wardana for a productive visit at the National Research Council.

\newpage
\begin{figure}
\caption{Exchange-correlation energy {\em vs.} filling factor for a spin-polarized
QHE system. The diamonds are results from numerical diagonalizations in the
two lowest Landau levels, and the solid line for $\nu>1$ is our analytical fit 
Eq. (\ref{eq:technical:an_appr}).}
\label{fig:num_energies}
\end{figure}
\begin{figure}
\caption{Local filling factors $\nu_\uparrow({\bf r})$, $\nu_\downarrow({\bf r})$, and
$\nu({\bf r})=\nu_\uparrow({\bf r})+\nu_\downarrow({\bf r})$ in the $v=0$ channel
for a 70 electron dot. The magnetic field strength is 3.5 T, $g^*=1.0$, 
and the edge width
is $w=8$ $l_B$. Here, the edge region is only partially polarized with a small
minority-spin density.}
\label{fig:v0_nu_1_dens}
\end{figure}
\begin{figure}
\caption{The local total filling factor and spin rotation angle $\theta(r)$ for
the $v=1$ channel of the same parameters as in Fig. \ref{fig:v0_nu_1_dens}. This
channel has a lower energy than the $v=0$ channel, and is locally completely polarized
with a non-trivial spin texture. That is, at any point $r$ the spin density is
polarized, but the spin direction is changing with position.}
\label{fig:v1_nu_1_dens}
\end{figure}
\begin{figure}
\caption{The local filling factor for the $v=0$ channel of a 40-electron
dot in the FQHE regime. The external magnetic field strength is 12 T, $g^*=1.0$, and
$w=4$ $l_B$.
In this case, the electron system is polarized.}
\label{fig:v0_nu_1/3_dens}
\end{figure}
\begin{figure}
\caption{The $v=1$ channel for the same system as in Fig. \ref{fig:v0_nu_1/3_dens}. This
channel is also completely polarized, but has a non-trivial spin-texture and lower energy
than the $v=0$ channel.}
\label{fig:v1_nu_1/3_dens}
\end{figure}
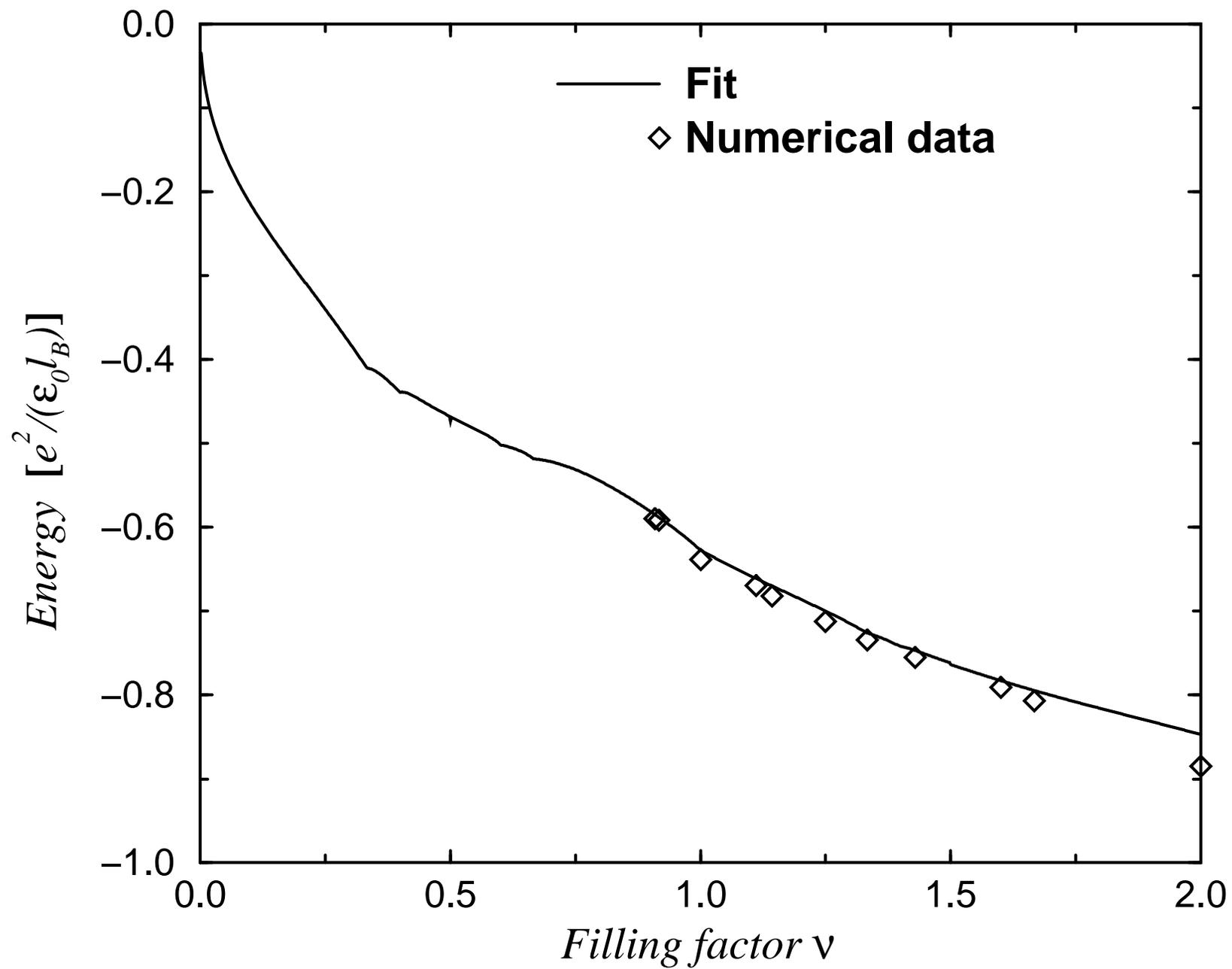
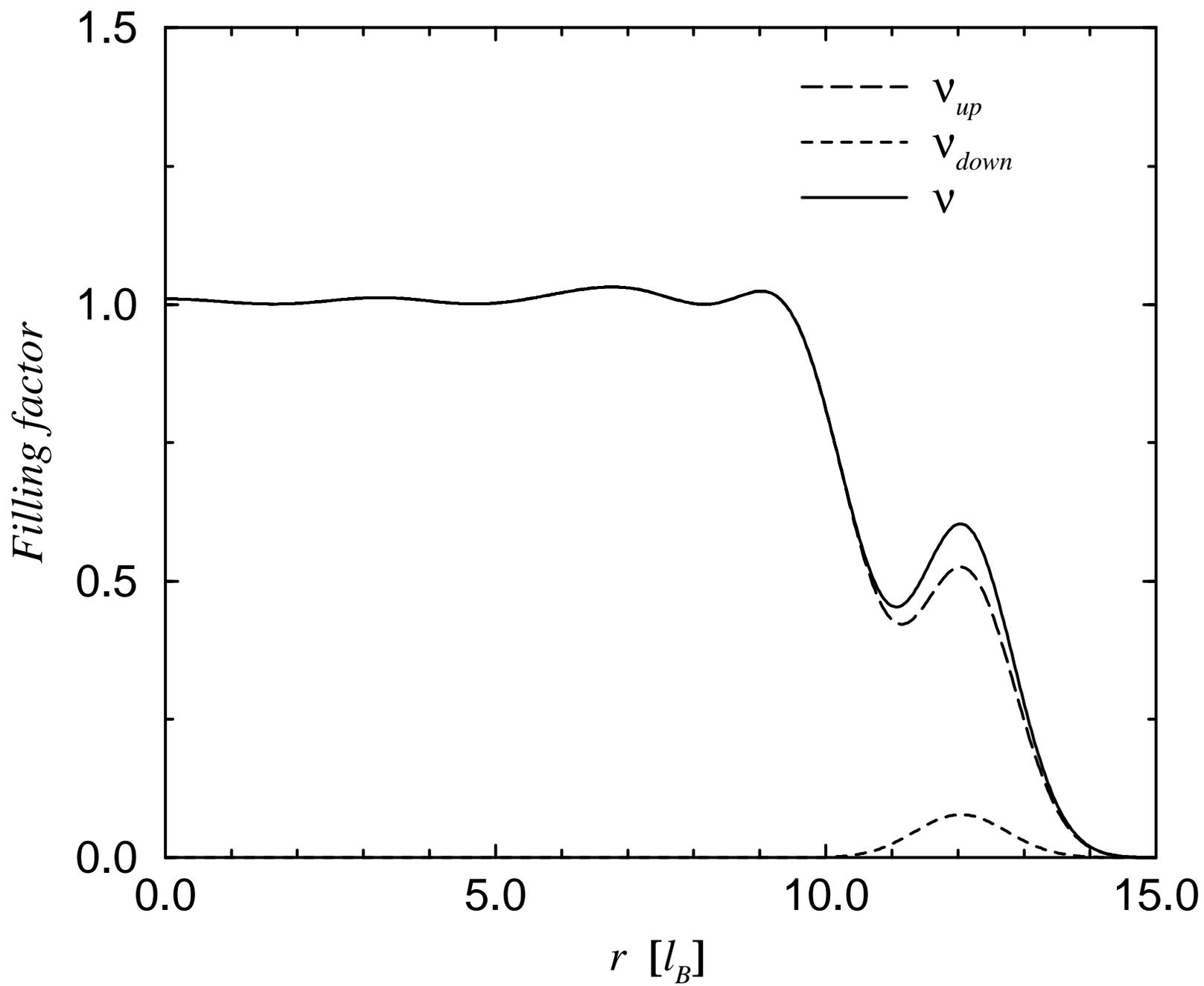
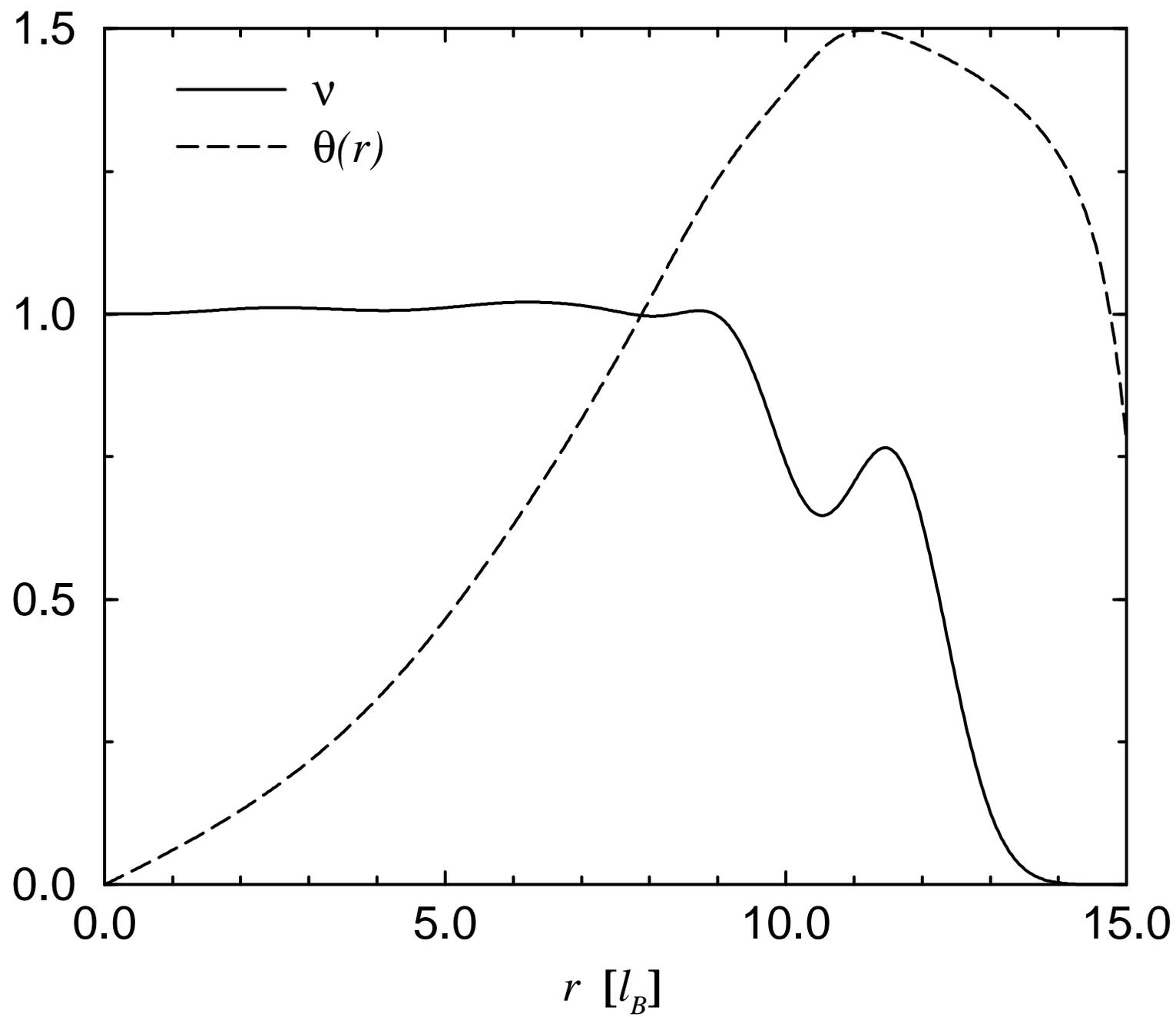
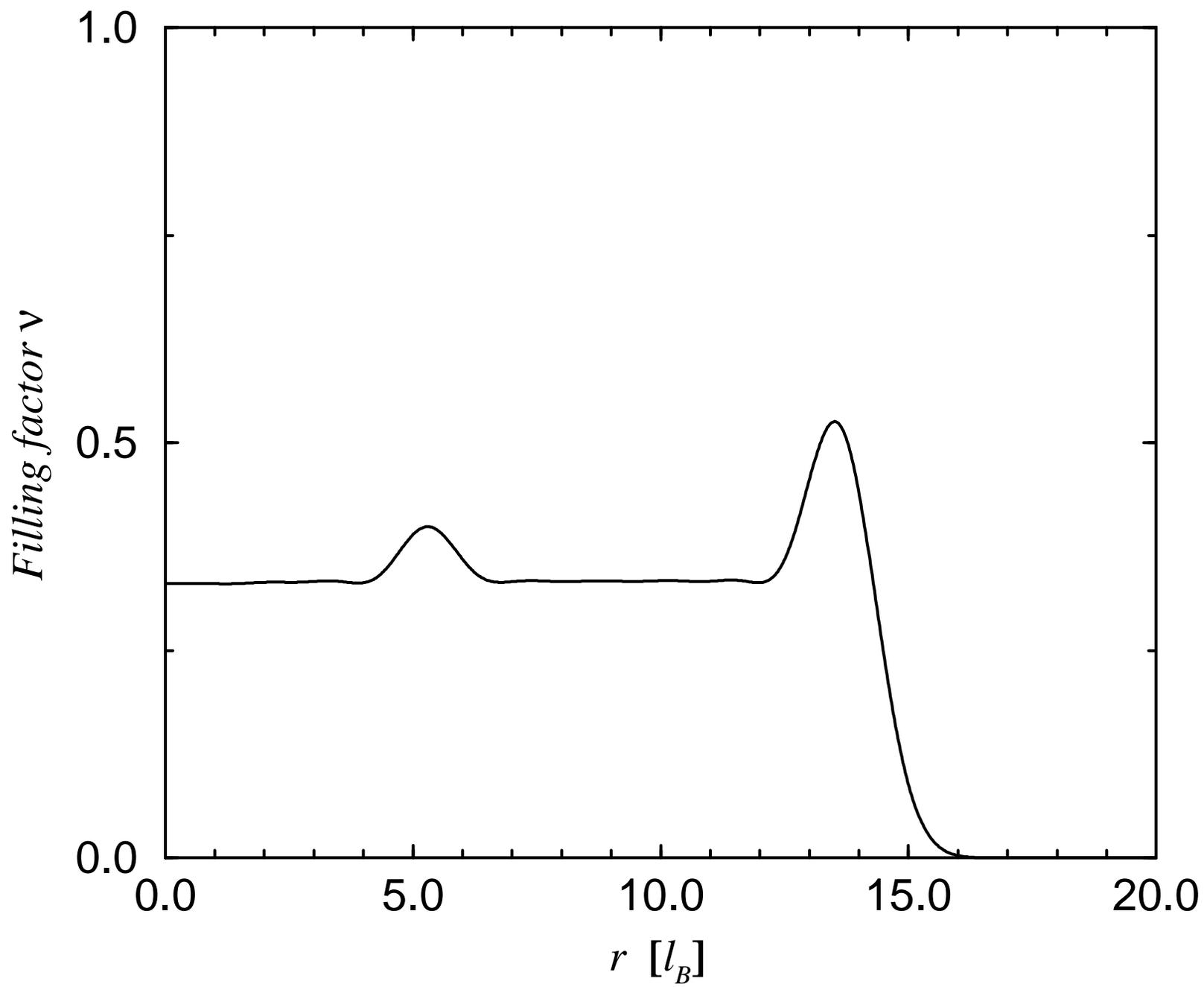
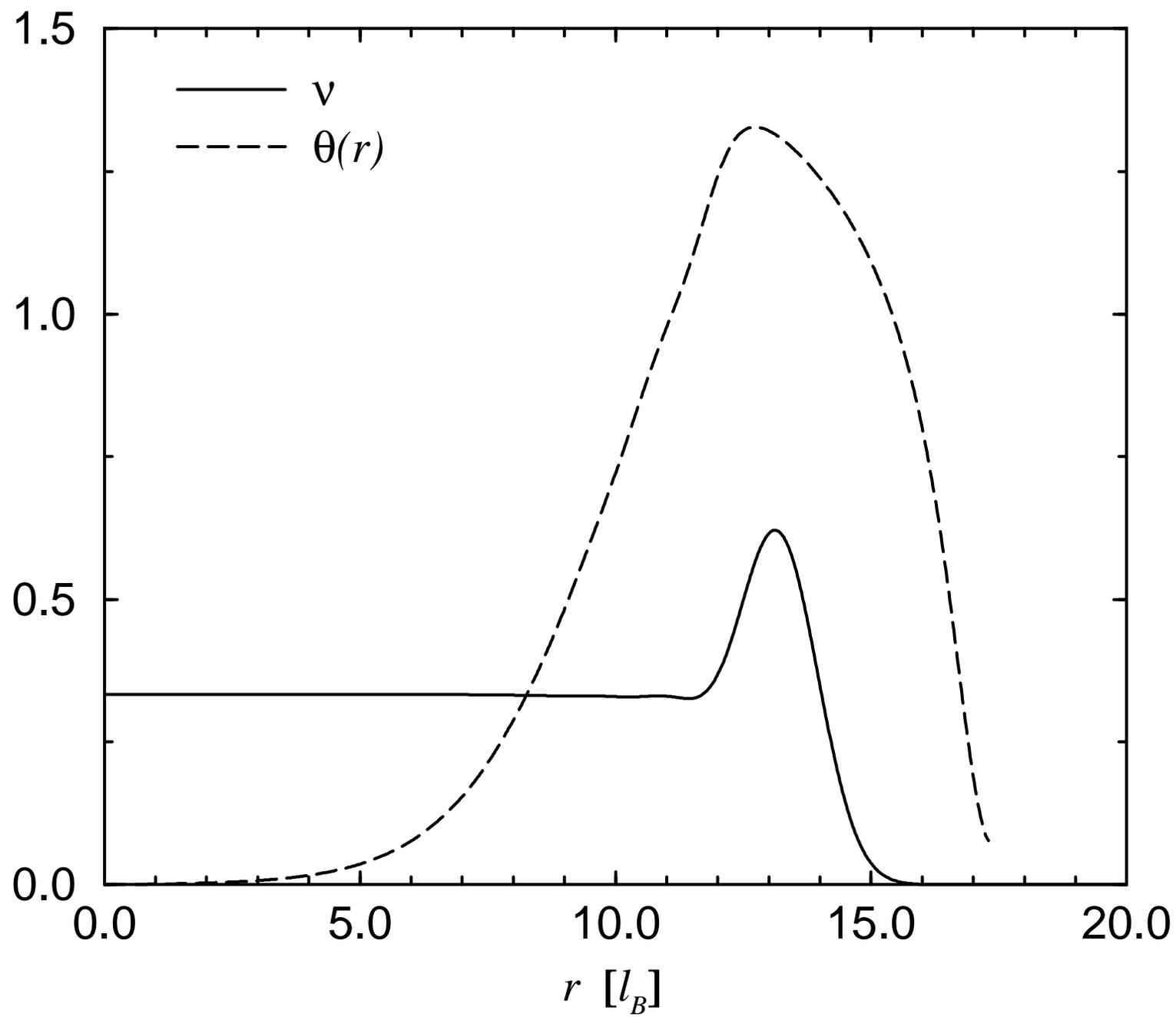
\end{document}